\documentclass[useAMS,usenatbib]{mn2e}

\usepackage{graphicx}
\usepackage{amsmath, amssymb}
\usepackage{array}

\begin{document}

\newcommand{\be}{\begin{equation}}
\newcommand{\ee}{\end{equation}}
\newcommand{\pdf}{\mathcal{P}}
\newcommand{\data}{d}
\newcommand{\Cb}{\mathcal{C}_b}
\newcommand{\mdl}{\mathcal{M}}
\newcommand{\lsim}{\,\raise 0.4ex\hbox{$<$}\kern -0.8em\lower 0.62ex\hbox{$\sim$}\,}
\newcommand{\gsim}{\,\raise 0.4ex\hbox{$>$}\kern -0.7em\lower 0.62ex\hbox{$\sim$}\,}

\newcommand{\params}{\theta}
\newcommand{\paramsU}{\boldsymbol{\theta}_\star}
\newcommand{\fsel}{f_\text{selec}}
\newcommand{\fobs}{f_\text{obs}}
\newcommand{\gal}{\text{gal}}
\newcommand{\stars}{\text{stars}}
\newcommand{\plan}{\text{planets}}
\newcommand{\mpl}{m_{Pl}}

\newcommand{\mean}{\boldsymbol{\mu}}
\newcommand{\like}{L}
\newcommand{\lnlike}{\mathcal{L}}
\newcommand{\ML}{^*}
\newcommand{\dr}{\textrm{d}}
\newcommand{\ie}{i.e.}
\newcommand{\eg}{e.g.}
\newcommand{\reion}{\text{re}}

\newcommand{\cd}{\cdot}
\newcommand{\cds}{\cdots}
\newcommand{\ip}{\int_0^{2\pi}}
\newcommand{\al}{\alpha}
\newcommand{\ba}{\beta}
\newcommand{\de}{\delta}
\newcommand{\De}{\Delta}
\newcommand{\ep}{\epsilon}
\newcommand{\Ga}{\Gamma}
\newcommand{\ka}{\tau}
\newcommand{\io}{\iota}
\newcommand{\La}{\Lambda}
\newcommand{\Om}{\Omega}
\newcommand{\om}{\omega}
\newcommand{\si}{\sigma}
\newcommand{\Si}{\Sigma}
\newcommand{\te}{\theta}
\newcommand{\ze}{\zeta}
\newcommand{\vth}{\ensuremath{\vartheta}}
\newcommand{\vph}{\ensuremath{\varphi}}
\newcommand{\MM}{\mbox{$\cal M$}}
\newcommand{\tr}{\mbox{tr}}
\newcommand{\hor}{\mbox{hor}}
\newcommand{\grad}{\mbox{grad}}
\newcommand{\cx}{\ensuremath{\mathbf{\nabla}}}
\newcommand{\lap}{\triangle}
\newcommand{\arctg}{\mbox{arctg}}
\newcommand{\bm}[1]{\mbox{\boldmath $#1$}}
\newcommand{\eff}{{\rm eff}}
\newcommand{\tto}{\Rightarrow}
\newcommand{\lag}{\langle}
\newcommand{\rag}{\rangle}
\newcommand{\fiso}{f_{\text{ci}}}
\newcommand{\Afiso}{\vert f_{\text{iso}}\vert}

\newcommand{\CI}{S_{c}}
\newcommand{\ND}{S_\nu}
\newcommand{\NV}{{V}_\nu}
\newcommand{\fnd}{f_{\text{ne}}}
\newcommand{\fnv}{f_{\text{nv}}}
\newcommand{\Df}{\Delta f}

\title[The isocurvature fraction]{The isocurvature fraction after WMAP 3--year data}

\author[Roberto Trotta]{Roberto Trotta\thanks{E-mail address: {\tt
rxt@astro.ox.ac.uk}}
\\
Oxford University, Astrophysics,  Denys Wilkinson Building, Keble
Road, OX1 3RH, United Kingdom}

\maketitle

\begin{abstract}
I revisit the question of the adiabaticity of initial conditions
for cosmological perturbations in view of the 3--year WMAP data. I
focus on the simplest alternative to purely adiabatic conditions,
namely a superposition of the adiabatic mode and one of the 3
possible isocurvature modes, with the same spectral index as the
adiabatic component.

I discuss findings in terms of posterior bounds on the
isocurvature fraction and Bayesian model selection. The Bayes
factor (models likelihood ratio) and the effective Bayesian
complexity are computed for several prior ranges for the
isocurvature content. I find that the CDM isocurvature fraction is
now constrained to be less than about $10\%$, while the fraction
in either the neutrino entropy or velocity mode is below about
$20\%$. Model comparison strongly disfavours mixed models that
allow for isocurvature fractions larger than unity, while current
data do not allow to distinguish between a purely adiabatic model
and models with a moderate (\ie, below about $10\%$) isocurvature
contribution.

The conclusion is that purely adiabatic conditions are strongly
favoured from a model selection perspective. This is expected to
apply in even stronger terms to more complicated superpositions of
isocurvature contributions.
\end{abstract}



\section{Introduction}

The detailed nature of the initial conditions for cosmological
perturbations is one of the open questions in cosmology. The
exquisite precision of the WMAP measurement of the first acoustic
peak location in the cosmic microwave background (CMB) temperature
power spectrum ($\ell = 220.7 \pm 0.7$, see \cite{Hinshaw:2006ia})
is a strong indication in favour of adiabatic initial conditions,
which predict for the first peak $\ell \approx 220$. The
alternative possibility of cold dark matter (CDM) isocurvature
initial conditions excites a sine wave (rather than the cosine
excited by adiabatic conditions) in the photon--baryon plasma,
resulting in a first acoustic peak displaced by half a period to
$\ell \approx 330$, see e.g.\ \cite{Trotta:2004qj,Durrer:2004fx}.
Furthermore, the ratio of the Sachs--Wolfe plateau for $\ell \lsim
50$ to the height of the peak is very different for the two modes.

A few years ago, Bucher and collaborators introduced two new
isocurvature modes, called ``neutrino density'' (or, more
appropriately, ``neutrino entropy'') and ``neutrino velocity''
modes \citep{Bucher:1999re}. They are characterized by a
non--vanishing initial entropy perturbation in the neutrino sector
and by a non--vanishing difference in the neutrino to photon
velocity, respectively. A superposition of the adiabatic and the
three isocurvature modes (cold dark matter, neutrino entropy and
neutrino velocity) constitutes the most general initial conditions
for the perturbations, at least if the Universe is radiation
dominated in its early phase \citep{Trotta:2004qj}. A baryon
isocurvature mode is observationally indistinguishable from a CDM
isocurvature one \citep{Bucher:2000hy,Gordon:2002gv} and can thus
be neglected without loss of generality.

Allowing for the most general type of initial conditions has two
effects on cosmological parameter extraction from CMB
measurements. First, the extra parameters associated with the
initial conditions introduce severe degeneracies which limit our
ability to reconstruct the cosmology
\citep{Trotta:2001yw,Trotta:2002iz}, even though this can
fortunately be remedied by using polarization information
\citep{Bucher:2000hy,Trotta:2004sm}. Secondly, it becomes
difficult to constrain the type of initial conditions, i.e.\ the
amount of isocurvature contributions allowed on top of the
predominantly adiabatic mode \citep{Moodley:2004nz}.

Recent works have investigated general isocurvature admixtures in
the initial conditions
\citep{Beltran:2005gr,Moodley:2004nz,Bean:2006qz}. In this work I
focus on the simplest alternative to a purely adiabatic power
spectrum, namely a superposition with one totally
(anti--)correlated isocurvature mode at the time with the same
spectral index as the adiabatic one. This is partly motivated by
models for the generation of initial conditions such as the
curvaton (see \eg~\cite{Gordon:2002gv,Lyth:2001nq} and references
therein), where this kind of scenario arises as a generic
prediction. A second justification comes from the model selection
approach used in the second part of this work. In comparing the
simplest (\ie, purely adiabatic) scenario with a more complex one,
it makes sense to start by adding a minimal number of extra
parameters, and see whether the extended model is justified by the
data. This model selection perspective has been recently advocated
by \cite{Beltran:2005xd,Trotta:2005ar}.

This paper is organized as follows: in section \ref{sec:isoc} we
introduce the parameterization of the initial condition parameters
space we are considering, while in section \ref{sec:bayes} we
review some concepts of Bayesian statistics and in particular the
model selection approach. We present our results in terms of
parameters constraint and model comparison outcome in section
\ref{sec:res} and offer our conclusions in \ref{sec:conc}.

\section{The isocurvature fraction }
\label{sec:isoc}

 The most general initial conditions for
cosmological perturbations are described by a symmetric $4\times4$
matrix, $M$, with $10$ free parameters representing the amplitudes
of the pure modes (along the diagonal) and their correlations
(off--diagonal elements). From a phenomenological point of view,
there are also $10$ more parameters describing the spectral tilt
of each mode and correlator. If one is willing to consider running
of the spectral index, then this would introduce another 10 free
parameters in the problem. As motivated in the introduction, we
consider here a minimal extension of the simplest adiabatic model,
namely a diagonal matrix
 \begin{align}
 M & = \text{diag}(\zeta, \CI, \ND, \NV) \\
   & = \zeta \cdot  \text{diag}(1, \fiso, \fnd,
 \fnv) , \label{eq:fx}
 \end{align}
where $\zeta$ is the amplitude of the curvature perturbation
(adiabatic mode), $\CI, \ND$ are the (gauge invariant) entropy
perturbations in the CDM and neutrino component defining
non--vanishing CDM isocurvature and neutrino entropy modes,
respectively. The neutrino entropy mode is often referred to as
``neutrino density''. The quantity $\NV$ corresponds to a
non--zero neutrino--photon velocity giving rise to a neutrino
velocity mode (see \eg~\cite{Trotta:2004qj} for precise
definitions). The quantities $\fiso, \fnd, \fnv$ give the
isocurvature fractions with respect to the curvature perturbation,
where the notation employed is ci = CDM isocurvature, ne =
neutrino entropy and nv = neutrino velocity. The sign of $f_x$
(with $x = $ ci, ne, nv) determines the nature of the correlation:
a positive correlation ($f_x > 0$) results in extra power to the
Sachs--Wolfe plateau, a negative correlation ($f_x < 0$) subtracts
power in the region $\ell \lsim 50$. As already mentioned, we take
a common spectral index for the adiabatic and the isocurvature
mode, $n_s$, and we analyse separately the three scenarios where
only one of the isocurvature modes is non--zero, in addition to
the adiabatic mode.

An alternative parameterization for the isocurvature contribution
that is common in the literature is given in terms of the variable
$\alpha_x$, or $\alpha_x^2$ (used \eg~by \cite{Beltran:2004uv} and
\cite{Bean:2006qz}). This is related to $f_x$ by
 \be \label{eq:alx}
 \alpha^2_x = \text{sign}(f_x)\frac{f_x^2}{1+f_x^2}.
 \ee
From a phenomenological perspective, there is little reason to
prefer one parameterization over the other. However, from a model
selection point of view the choice of the variable one puts flat
priors on is of great importance, since the available parameter
space under the prior enters in the calculation of the Occam's
factor for the model, see the discussion in \cite{Trotta:2005ar}.
We must consider the choice of priors as inherent to the
specification of the extended model and different choices will
lead to different conclusions since the Occam's razor effect is
not invariant under non--linear transformations of variables.

Once a fundamental model for the generation of the initial
condition is specified, one can select the appropriate physical
variable over which to impose a prior reflecting our state of
knowledge before we see the data. For instance, it can be argued
that the $f_x$ parameterization is closer to the curvaton setup,
while the $\al_x$ choice of variable compresses the parameter
space in the compact interval $-1 < \al_x < 1$. A flat prior of
$\al_x$ gives equal {\em a priori} accessible volume to
adiabatic--dominated ($|\al_x| \leq 0.5$) and to
isocurvature--dominated ($|\al_x| > 0.5$) models.  The prior on
$f_x$ is very much dependent on what we think the available
parameter space is under our extended model. Therefore we discuss
below the results of model selection as a function of the prior
width $\Df$, taking a flat prior in the range $-\Df \leq f_x \leq
\Df$. This allows an easy comparison once a prior range under a
specific model is given. We postpone to a future work a detailed
analysis of prior selection based on first principles.

\section{Parameter estimation, model selection and model complexity} \label{sec:bayes}

Bounds on the isocurvature fraction are derived in terms of high
probability regions in the posterior probability density function
(pdf) for the parameters $\params$ given the data $\data$,
$p(\params | \data)$. This is obtained through Bayes theorem,
 \be p(\params | \data) = \frac{p(\data | \params)
 \pi(\params)}{p(\data)}, \label{eq:bayes}
\ee
 where $p(\data | \params)$ is the likelihood function,
$\pi(\params)$ is the prior pdf and $p(\data)$ is the model
likelihood (sometimes called ``the evidence'') of the data under
the model. The model under consideration is defined by the
parameter set $\params$ {\it and} the choice of the prior
$\pi(\params)$ (we shall return to this point below).

The model likelihood is a normalization constant independent on
the parameters of the model, and it can be ignored as far as the
parameter estimation step is concerned. It becomes the key
quantity for model selection, and in particular we are interested
in the relative odds between the simplest, purely adiabatic model
$M_0$ and a model augmented by an extra isocurvature contribution,
$I_x = (M_0, f_x)$, with $x =$ ci, ne, nv as above. The change in
our degree of belief in the two models after we have seen the data
is described by the {\em Bayes factor}
 \be
 B = \frac{p(\data|M_0)}{p(\data|I_x)},
 \ee
which is the ratio of the normalization constants for the two
models in Bayes theorem, Eq.~\eqref{eq:bayes}. Since the two
models are nested (\ie, we obtain $M_0$ from $I_x$ by setting the
isocurvature fraction to zero, $f_x = 0$), the Bayes factor can be
conveniently computed using the Savage--Dickey density ratio
(SDDR) (see \cite{Trotta:2005ar} and references therein)
 \be \label{eq:bayesf}
 B = \left. \frac{p(f_x |\data, I_x)}{\pi(f_x  | I_x)}\right|_{f_x = 0}.
 \ee
This is easy to compute from a Monte Carlo Markov Chain (MCMC),
requiring only knowledge of the properly normalized posterior over
the extra variable $f_x$ of the extended model. Furthermore, using
the SDDR has the advantage that the impact of a change of prior
can usually be evaluated by simply post--processing a chain
including the new prior. This is the approach used below in
section~\ref{sec:res}. If the posterior pdf is well approximated
by a Gaussian distribution with mean $\mu$ and standard deviation
$\sigma$, and for a flat prior in the range $-\Df \leq f_x \leq
\Df$, the Bayes factor \eqref{eq:bayesf} becomes
 \be \label{eq:Gauss}
 B = \sqrt{\frac{8}{\pi}}\frac{\Df}{\sigma}
 \left[\text{erfc}\left(-\frac{|\mu| - \Df}{\sqrt{2}\sigma}\right)
- \text{erfc}\left(\frac{|\mu| + \Df}{\sqrt{2}\sigma}\right)
  \right]^{-1}.
 \ee

 For $\ln B> 0$ model $M_0$ is favoured over $I_x$ because the
extra complexity (in terms of wasted parameter space) of $I_x$ is
not warranted by the data, while for $\ln B < 0$ $I_x$ is favoured
since the data require the extra parameter. A useful rule of
thumb~\citep{Kass} is that a positive (strong) preference requires
$|\ln B| \gsim 1 \,(\gsim 3)$. A model likelihood ratio $|\ln B |
> 5$ (corresponding to odds $>150:1$ is deemed to constitute ``decisive''
evidence.)

Finally, the last relevant quantity for our analysis is the
Bayesian model complexity, which measures the number of parameters
the data can support, regardless of whether the parameters in
question are actually detected or not (for more details, see
\cite{Kunz:2006mc}). The {\em Bayesian complexity} is defined as
 \be
 \Cb = \overline{\chi^2(\params)} - \chi^2(\hat{\params}),
 \ee
where the effective $\chi^2(\params)$ is derived from the
likelihood as $\chi^2(\params) = -2 \ln p(\data|\params)$. The bar
denotes an average over the posterior pdf, while the hat denotes a
point--estimator which in our case we take to be the mean under
the posterior, \i.e. $\hat{\params} = \overline{\params}$. We will
use $\Cb$ to quantify the number of supported parameters in our
extended models $I_x$, in order to verify whether the isocurvature
fraction is a variable that could have been detected using current
data. A detailed discussion of the meaning and interpretation of
the Bayesian complexity can be found in \cite{Kunz:2006mc}.

It is important to stress that both the model likelihood and the
Bayesian complexity depend not only on the data but also on the
model description one chooses to adopt, \ie\ on the prior choices
one makes for $\pi(f_x)$ (see \cite{Trotta:2006pw} for an example
applied to the case of dark energy models). This is an irreducible
feature of the Bayesian model selection approach. It seems to us
that there cannot be an absolute notion of ``a best model'', but
only relative statements about the support the data give to
different models when compared to each other. Furthermore, the
very concept of Bayesian complexity is only meaningful when the
constraining power of the data is compared to the scale of the
problem at hand, which again must be defined by specifying the
prior.

Our simplest model $M_0$ is a flat $\Lambda$CDM Universe with
purely adiabatic conditions, described by following set of 6
parameters
 \be
 \params = (\zeta, \om_b, \om_c, \Theta_\star, n_s, \tau_r)
  \ee
where $\zeta$ is the curvature perturbation, $\om_b, \om_c$ are
the physical densities of baryons and CDM, respectively,
$\Theta_\star$ is the ratio of the sound horizon to the angular
diameter distance to last scattering, $n_s$ is the spectral tilt
and $\tau_r$ the optical depth to reionization. An extra bias
parameter $b$ for the matter power spectrum is treated as a
nuisance parameter and marginalized over, hence we do not count it
as an additional parameter. We do not consider tensor modes nor
extra neutrino species nor running of the spectral index. We take
our 3 neutrino families to be massless and we fix the dark energy
equation of state to $w=-1$ at all redshifts. All of those choices
are motivated by the fact that inclusion of any of the above extra
parameters is presently not required by the data. This means that
a comparison between a model including both the isocurvature
fraction and one of the above extra parameters against the simple
adiabatic model would favour even more strongly the latter, as a
consequence of the extra Occam's factor effect coming from the
extra parameter. In this sense, our model selection is actually
conservative.

The situation is different for parameter constraints, since in
this case strong degeneracies between the isocurvature fraction
and other extra parameters might change the posterior bounds of
$f_x$. In particular, one can expect a strong degeneracy between
the CDM isocurvature mode and the presence of a tensor mode from
gravity waves, when considering temperature power spectrum
information alone. The extra power contributed by the CDM
isocurvature to the Sachs--Wolfe plateau for small $\ell$'s is
strongly anti--correlated with the tensor mode amplitude. However,
inclusion of polarization data would help in breaking this
degeneracy, at least partially. The impact of allowing a tensor
mode contribution is very mild for the neutrino modes, since the
Sachs--Wolfe plateau is lower than the first acoustic peak for
these modes, and as a consequence constraints on their amplitudes
are dominated by the height of the peak, not by the height of the
plateau. Similar considerations also apply for a possible running
of the spectral index. Other parameters that mainly affect the
angular diameter distance to last scattering and therefore the
position of the acoustic peaks in the spectrum (such as the dark
energy equation of state, the curvature of spatial sections or an
extra background of relativistic particles) present only weak
degeneracies with the isocurvature fractions, since the peaks'
position is strongly constrained by the data.

In the following, we therefore limit our analysis to the 6
parameters model $M_0$ described above, complemented in the
extended models by the isocurvature fractions as follows. The
extended models $I_x$ contain a non--vanishing isocurvature
fraction
 \be
 I_x = (\params, f_x)
 \ee
where $f_x$  is defined in Eq.~\eqref{eq:fx} and $x =$ ci, ne, nd.
The spectral index of the isocurvature mode is the same as the
adiabatic one, $n_s$. The correlation coefficient between the
adiabatic and the isocurvature mode is $\pm 1$, depending on the
sign of $f_x$.

\section{Results and discussion} \label{sec:res}

In this section we present our results about the isocurvature
fraction in terms of posterior bounds, Bayesian model selection
and effective model complexity.

We use the WMAP 3--year temperature and polarization data
\citep{Hinshaw:2006ia,Page:2006hz} supplemented by small--scale
CMB measurements \citep{Readhead:2004gy,Kuo:2002ua}. We add the
Hubble Space Telescope measurement of the Hubble constant $H_0 =
72 \pm 8$ km/s/Mpc \citep{Freedman:2000cf} and the Sloan Digital
Sky Survey (SDSS) data on the matter power spectrum on linear
scales \citep{Tegmark:2003uf}.

\begin{figure}
\centering
\includegraphics[width=\linewidth]{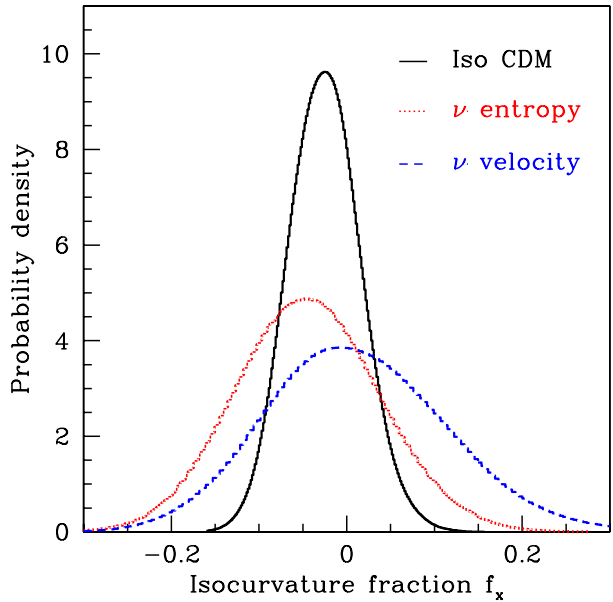}
\caption[My figure]{Normalized posterior probability density for
the isocurvature fraction parameter $f_x$. CMB and large scale
structure data are compatible with purely adiabatic initial
conditions, with a slight tendency towards negatively correlated
isocurvature components.} \label{fig:fiso}
\end{figure}

In Figure~\ref{fig:fiso} we plot the 1--dimensional, marginalized
posterior pdf on the isocurvature fraction parameter $f_x$. We we
have adopted a flat prior of $f_x$ of width much larger than the
posterior, so that the range of the prior does not influence the
result. The isocurvature fraction is compatible with zero for all
three isocurvature modes, with a slight shift of the peak of the
pdf to negative values. This corresponds to a negative
correlation, in which case the contribution to the large scales
CMB power due to the isocurvature auto--correlation spectrum is
largely compensated by the negative correlator. The posterior mean
and standard deviation for $f_x$ are given in
Table~\ref{tab:bounds}, as well as 1--dimensional marginalized
intervals encompassing $95\%$ of probability. We find that the
isocurvature fraction for the CDM mode is constrained to be $-0.10
< f_{\text{ci}} < 0.06$ ($95\%$ probability), while for the two
neutrino modes we obtain $-0.20 < f_{\text{ne}} < 0.12$ (neutrino
entropy) and $-0.18 < f_{\text{nv}} < 0.22$ (neutrino velocity).
We notice that the tightest constrained mode is the CDM
isocurvature one. This is because with our definition of $f_x$,
for a given value of $f_x$ the CDM isocurvature is the mode with
the largest contribution to the CMB power spectrum. Also, all of
the 1--dimensional pdf's for $f_x$ are very close to Gaussian.
Hence we expect that Eq.~\eqref{eq:Gauss} is a good approximation
to the Bayes factor, Eq.~\eqref{eq:bayesf}, as we now show.

\begin{table}
\caption{Constraints on the isocurvature fraction $f_x$, from
WMAP3 and other CMB data, and SDSS matter power spectrum
measurements. We give the posterior mean $\mu$ and standard
deviation $\sigma$, as well 1--dimensional marginalized regions
encompassing $95\%$ of posterior probability.} \label{tab:bounds}
\centering
\begin{tabular}{|l | ccc|}
\hline Model      & $\mu$              & $\sigma$            &  95\% interval on $f_x$   \\
\hline
CDM iso           & $-2.5\cdot10^{-2}$ & $4.0 \cdot 10^{-2}$ & $-0.10\dots0.06$  \\
$\nu$ entropy     & $-4.4\cdot10^{-2}$ & $8.0 \cdot 10^{-2}$ & $-0.20\dots0.12$   \\
$\nu$ velocity    & $-1.2\cdot10^{-2}$ & $1.0 \cdot 10^{-1}$ &
$-0.18\dots0.22$ \\\hline
\end{tabular}
\end{table}

\begin{figure}
\centering
\includegraphics[width=\linewidth]{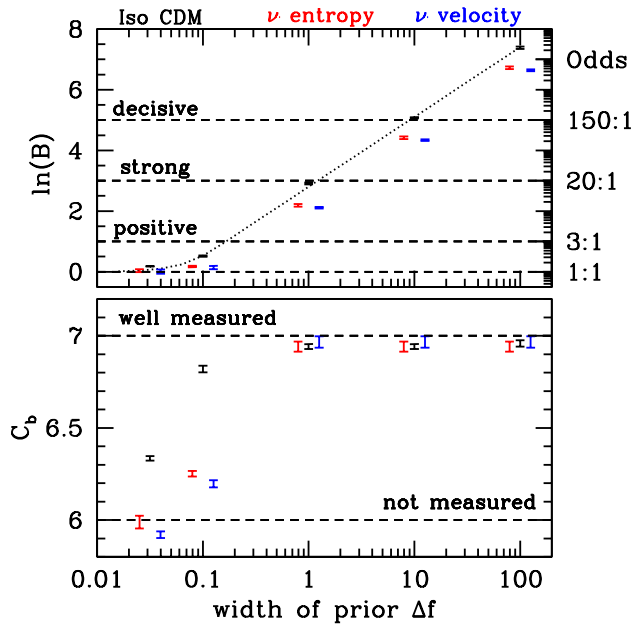}
\caption{Result of model selection between a purely adiabatic
model and an extended model featuring a totally (anti--)correlated
isocurvature component, as a function of the prior available range
for the isocurvature fraction, $\Df$. Top panel: the Bayes factor
strongly disfavours models with $\Df \gg 1$ because of the Occam's
razor effect, while models predicting an isocurvature fraction
below about $10\%$ in any of the three modes cannot presently be
ruled out ($\ln B < 1$). The dotted line gives the Gaussian
approximation to the Bayes factor, Eq.~\eqref{eq:Gauss}, for the
CDM isocurvature mode. Bottom panel: the Bayesian complexity gives
the effective number of parameters the data can support. For $\Df
\lsim 1$ the isocurvature component in the neutrino entropy and
velocity modes is not supported by the data. The errors have been
computed as the variance of 10 random sub--chains, and the
neutrino entropy and velocity modes have been shifted horizontally
to the left and to the right, respectively, for clarity.}
\label{fig:bayes}
\end{figure}

We now evaluate the Bayes factor between the models including an
isocurvature contribution and the simplest, purely adiabatic
model. As we have seen above in the parameter extraction step,
there is no indication that the data require and isocurvature
component, since the isocurvature fraction is compatible with 0 .
This is consistent with the findings of \cite{Bean:2006qz}. We
therefore expect the Bayes factor to favour the purely adiabatic
model on the ground of the Occam's razor argument. The strength of
evidence in favour of the adiabatic model depend on the amount of
wasted parameter space for the isocurvature fraction, \ie~by the
prior range $\Df$. In the top panel of Figure~\ref{fig:bayes}, we
plot the Bayes factor as a function of the prior width $\Df$,
while in the bottom panel we plot the Bayesian complexity, \ie~the
number of parameters effectively constrained by the data. We can
see that for models with poor predictivity, \ie~a large prior
accessible range $\Df \gg 1$, one finds strong ($>20:1$) to
decisive ($>150:1$) posterior odds against the extended model for
all of the three isocurvature modes. We also plot the Gaussian
approximation to the SDDR for the Bayes factor,
Eq~\eqref{eq:Gauss}, for the CDM isocurvature mode, and find a
very good match with the value computed numerically from the Monte
Carlo chain.

For a prior choice $\Df \leq 1$, the Bayesian complexity is close
to 7, indicating that all of the 7 parameters of the extended
model have been measured. We therefore conclude that models
predicting up to the same amount of isocurvature to adiabatic
power (the case $\Df = 1$) are strongly disfavoured for the CDM
mode, and mildly disfavoured in the case of the two neutrino
modes. However, if the prior range is reduced below $\Df = 1$,
\ie~for models predicting predominantly adiabatic initial
conditions with subdominant isocurvature contribution, the Bayes
factor gives an inconclusive result, with about equal odds for the
purely adiabatic and the mixed models. At the same time, the
Bayesian complexity decreases, indicating that $f_x$ is only
poorly constrained with respect to the scale of the prior,
especially for the neutrino density and velocity modes. This
reinforces the conclusion that current data are not strong enough
to select among a purely adiabatic model and one which predicts up
to $10\%$ isocurvature contribution and we need to acquire better
data in order to obtain a higher--odds result.

\section{Conclusions} \label{sec:conc}

We have submitted the question of the type of initial conditions
for cosmological perturbations to renewed scrutiny in the light of
WMAP 3--year data. We have focused on the simplest and well
motivated alternative to a purely adiabatic model, namely an
admixture of one totally (anti--)correlated isocurvature mode at
the time, with the same spectral tilt as the adiabatic one.

We have derived posterior bounds on the isocurvature fraction from
WMAP 3--year data combined with other CMB measurements and SDSS.
We have constrained the isocurvature fraction in the CDM mode to
be less than about $10\%$, while the maximum allowed neutrino
isocurvature contribution (either density or velocity) is about
$20\%$.

Bayesian model selection tends to favour purely adiabatic initial
conditions with strong odds ($>20:1$) when compared to models
predicting isocurvature fractions larger than unity. For such
models -- having a large prior range on the isocurvature fraction
-- we have shown that the data can support 7 parameters, but that
only 6 of them are required, with no need to include isocurvature
modes from a model selection point of view. These findings confirm
the conclusions of~\cite{Kunz:2006mc}. However, mixed models that
limit the isocurvature contribution to less than about $10\%$
cannot presently be ruled out. We have shown that the constraining
power of the data for this class of models is insufficient, and
therefore we must hold our judgement until better data becomes
available. These findings are however dependent on the
parameterization chosen for the isocurvature fraction, that in
this work is motivated by the curvaton scenario. The question of
prior selection will be further addressed in a future publication.

It is reasonable to expect that the same conclusion would apply in
even stronger terms to the case of more complicated models,
\eg~those involving a superposition of different isocurvature
modes at the same time, or with arbitrary correlations among them.
In fact, more complicated models (such as the class considered by
\cite{Bean:2006qz}) ought to be even more disfavoured because of
their larger volume of wasted parameter space. At present, Occam's
razor is perfectly compatible with the simplest possibility,
namely purely adiabatic initial conditions.

We think that this model comparison approach can be a useful
complement to parameter constraints analysis, and that it can
offer valuable guidance in building models for the generation of
primordial perturbations.

{\em Acknowledgments} R.T. is supported by the Royal Astronomical
Society through the Sir Norman Lockyer Fellowship. I am grateful
to Andrew Liddle and David Parkinson for comments. I acknowledge
the use of the package \texttt{cosmomc}, available from
\texttt{cosmologist.info} and the use of the Legacy Archive for
Microwave Background Data Analysis (LAMBDA). Support for LAMBDA is
provided by the NASA Office of Space Science.


\end{document}